\documentclass{webofc}
\usepackage[varg]{txfonts}   

\begin{document}
\title{Search for correlations of high-energy neutrinos and ultra-high-energy cosmic rays}

\author{\firstname{Lisa} \lastname{Schumacher}\inst{1}\fnsep\thanks{\email{lisa.schumacher@icecube.wisc.edu}} for the ANTARES, IceCube, Pierre Auger and Telescope Array collaborations
}
\institute{III. Physikalisches Institut B, RWTH Aachen University}
\abstract{%
The IceCube Neutrino Observatory has recently found compelling evidence for a particular blazar producing high-energy neutrinos and $\mathrm{PeV}$ cosmic rays, however the sources of cosmic rays above several $\mathrm{EeV}$ remain unidentified. It is believed that the same environments that accelerate ultra-high-energy cosmic rays (UHECRs) also produce high-energy neutrinos via hadronic interactions of lower-energy cosmic rays. Two out of three joint analyses of the IceCube Neutrino Observatory, the Pierre Auger Observatory and the Telescope Array yielded hints for a possible directional correlation of high-energy neutrinos and UHECRs. These hints however became less significant with more data. Recently, an improved analysis with an approach complementary to the other analyses has been developed. This analysis searches for neutrino point sources in the vicinity of UHECRs with search windows estimated from deflections by galactic magnetic fields. We present this new analysis method for searching common hadronic sources, additionally including neutrino data measured by ANTARES in order to increase the sensitivity to possible correlations in the Southern Hemisphere.
}
\maketitle
\section{Multimessenger introduction}
\label{sec:intro}
The origin of cosmic rays is an over century-old puzzle. 
While we have compelling evidence for an extra-galactic source of sub-$\mathrm{PeV}$ neutrinos and $\mathrm{PeV}$ cosmic rays (TXS 0506+056~\cite{ic:txs}), the sources of ultra-high-energy cosmic rays above $\mathrm{1 EeV}$ (UHECRs) are not yet identified.
The exact source positions are difficult to establish due to the deflection of the charged UHECRs in Galactic and intergalactic magnetic fields. 
This uncertainty adds to the unknown rigidity of the UHECRs.
Therefore, it is useful to combine the UHECR data with the information concerning neutrinos in a multimessenger approach.
Cosmic rays produce charged and neutral mesons, 
which ultimately decay into charged leptons and neutrinos, and photons, respectively.
However, only the high-energy neutrinos are definite tracers of the hadronic interactions of cosmic rays.
Although neutrinos in the $\mathrm{TeV}$ to $\mathrm{PeV}$ energy range do not originate directly from UHECR 
interactions,
a calorimetric accelerator environment that confines their lower-energy counterparts with energies $<\!100~\mathrm{PeV}$ can indeed produce neutrinos in the currently accessible energy range~\cite{theo:waxman_murase}.
We present two established methods~\cite{icrc:cr_nu_correlation} (A, B), a new method 
using the complementary high-statistics neutrino dataset 
(C), and their results for the search for spatial correlations of neutrinos measured by ANTARES (ANT) and IceCube (IC) and UHECRs measured by the Pierre Auger Observatory (PA) and Telescope Array (TA).

\section{The data sets}
\label{sec:obs}
\textbf{The ANTARES data set} used for analysis C consists of 7622 muon neutrino candidates detected between 2007 and 2015, and has been optimized for point source searches.
It has a median angular difference between the reconstructed muon direction and the parent neutrino of $0.4^\circ$. 
The reconstructed $\sin \delta$  of the events lies between $-1$, i.e. South Pole, and $+0.8$, with the bulk of events originating from the Southern Hemisphere. 
This data set has been used in ANT's latest all-flavor point-source analysis~\cite{ant:9yrps}.

\textbf{The IceCube data sets} are two different full-sky samples. 
The first data set consists of 1.2 million muon neutrino candidates from both hemispheres recorded between 2008 and 2017, and has been optimized for point-source searches as required for analysis C.
The median of the angular difference between the reconstructed muon direction and the parent neutrino is $0.5^\circ$ in the energy range most sensitive to point-source searches. 
This data set has been used in the time-dependent point-source analysis at the position of TXS 0506+056~\cite{ic:txs}.
The second data set consists of 58+15 high-energy neutrino candidates measured in the shower channel (angular resolution $\sim\!15^\circ$) and in the track channel (angular resolution $<\!1^\circ$), respectively, between 2010 and 2016~\cite{icrc:hese2017} and is optimized for a pure astrophysical neutrino signal as required for analyses A and B.
The data set is completed with 35 through-going highest-energy muon neutrino candidates with energies $E>\!200~\mathrm{TeV}$, equivalent to seven years data from the astrophysical diffuse flux analysis~\cite{haack:2017ICRCdiff}.

\textbf{The Pierre-Auger data set} consists of 231 UHECR events with reconstructed energies $>\!52~\mathrm{EeV}$ recorded with the surface detector array between 
2004 and 
2014.
The angular resolution is $<\!0.9^\circ$, and the field of view covers the Southern Hemisphere and the Northern Hemisphere up to $45^\circ$.
This data set has been used in PA's anisotropy analyses~\cite{auger:anisotropies}.

\textbf{The Telescope-Array data set} consists of 109 UHECR events with reconstructed energies $>\!57~\mathrm{EeV}$ recorded with the scintillator array between 2008 and 2015.
The angular resolution is $<\!1.5^\circ$, and the field of view covers mainly the Northern Hemisphere.
Six years of this data set have been used in TA's hotspot analysis with one additional year being supplied for this work~\cite{ta:anisotropy_hotspot}.
In order to match the energy spectrum measured by PA and TA in the ankle region, the energies of the TA events have been scaled down by $13\%$~\cite{original-uhecr-nu}.

\section{Overview over analysis methods and previous results}
\label{sec:method}

\textbf{The cross-correlation method} (analysis A) is based on counting the number of UHECR-neutrino pairs
as a function of the maximal angular separation. 
The experimentally computed values are then compared to the averaged background, which is calculated once for an isotropic neutrino flux and once for an isotropic UHECR flux. 
This analysis is performed on the combined UHECR data set and on the track-like and shower-like high-energy neutrino candidates separately.
The smallest post-trial p-values found are $5.4\cdot 10^{-3}$ and $1.0\cdot 10^ {-2} $ within angular distances of $22^\circ$ and $16^\circ$ calculated for an isotropic flux of UHECR and neutrinos, respectively, for cascade-like events~\cite{icrc:cr_nu_correlation}.
These p-values are less significant compared to the first analysis reported in~\cite{original-uhecr-nu}, 
where p-values of $5.0\cdot 10^{-4}$ and $8.5\cdot10^{-3}$ were found.

\textbf{The unbinned likelihood method} (analysis B) is based on a stacking point-source analysis applied to the UHECRs, where the neutrino arrival directions are smeared with a symmetric 2D Gaussian and stacked as the signal template.
The width of the Gaussian is the quadratic sum of UHECR reconstruction uncertainty and of the magnetic deflection determined by $\sigma_{MD}=D/E_{CR}[100~\mathrm{EeV}],~D\in[3^\circ, 6^\circ, 9^\circ]$. 
This analysis is applied to the combined UHECR data set and to track-like and cascade-like high-energy neutrino candidates separately.
The smallest post-trial p-values found are $2.2\cdot 10^{-2} $ and $ 1.7\cdot 10^ {-2} $ for the magnetic deflection parameter $D=6^{\circ}$, which again occurred for cascade-like events for an isotropic flux of UHECR and neutrinos, respectively~\cite{icrc:cr_nu_correlation}.
These p-values are also less significant compared to the first result obtained in~\cite{original-uhecr-nu}, 
where p-values of $8.0\cdot 10^{-4}$ and $1.3\cdot10^{-3}$ were found.

\section{New combined UHECR-neutrino likelihood}
\label{sec:myllh}

\subsection{Search method}
\label{sec:mymethod}
This analysis (C) is based on an unbinned likelihood method for searching point-like neutrino sources~\cite{ic:ps_7yr}, where the number of neutrino signal events $n_s$ and the spectral index $\gamma_s$ are optimized on grid positions $\overrightarrow{x_s}$ covering the whole sky. 
The result is a test statistic map of the neutrino sky defined as 
$TS(\overrightarrow{x_s}) = -2\log \mathcal L\left ( \hat{n}_s, \hat{\gamma}_s \right )/\mathcal L \left (n_s=0 \right ) $.
The signal hypothesis is that point-like neutrino sources are spatially correlated with UHECR arrival directions, subject to a specific magnetic deflection hypothesis.
One UHECR arrival direction and the corresponding magnetic smearing $\sigma_{MD}$ are used to construct a 2D Gaussian which is logarithmically added to the test statistic map.
The result is an effective selection of the neutrino sky where the largest remaining test statistic spot is the most likely neutrino source counterpart.
This is equivalent to optimizing $n_s$, $\gamma_s$ and the source position in presence of a spatial, Gaussian prior.
This procedure is repeated for all selected UHECRs, and the resulting test statistic values are summed to yield the final test statistic.

\subsection{Results}
\label{sec:results}
\begin{figure} [!b]
  \centering
  \sidecaption
  \includegraphics[width=0.7\textwidth]{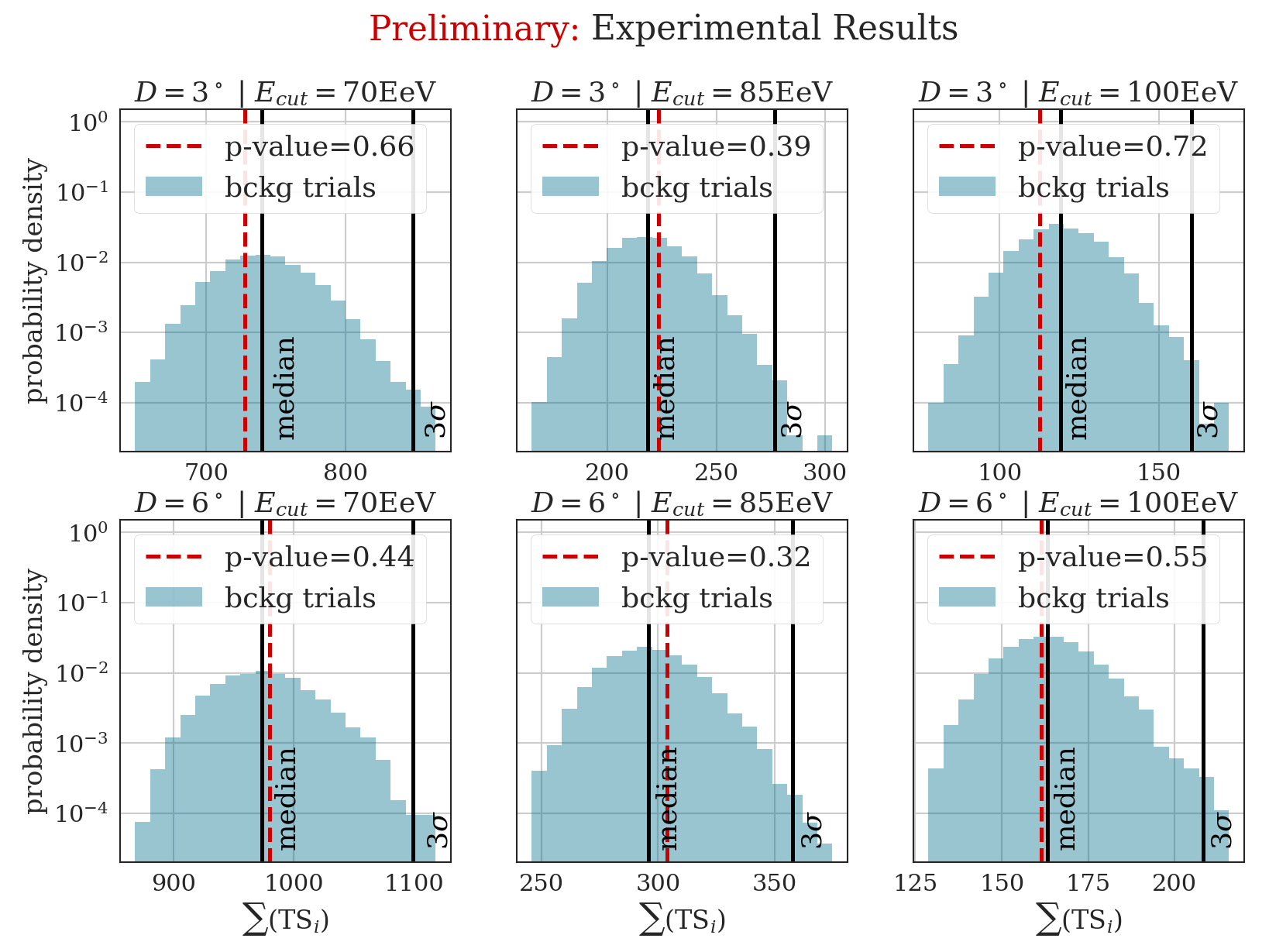}
  \caption{Experimental results (red dashed line) w.r.t. the corresponding background test statistic distribution (blue histogram), where the background median and the $3\sigma$ quantile have been marked (black lines). The magnetic deflection parameter $D$ and the energy cut on the UHECRs are given in the titles.}
  \label{fig:ts_result}
\end{figure}

\begin{figure}
  \centering
  \sidecaption
  \includegraphics[width=\textwidth]{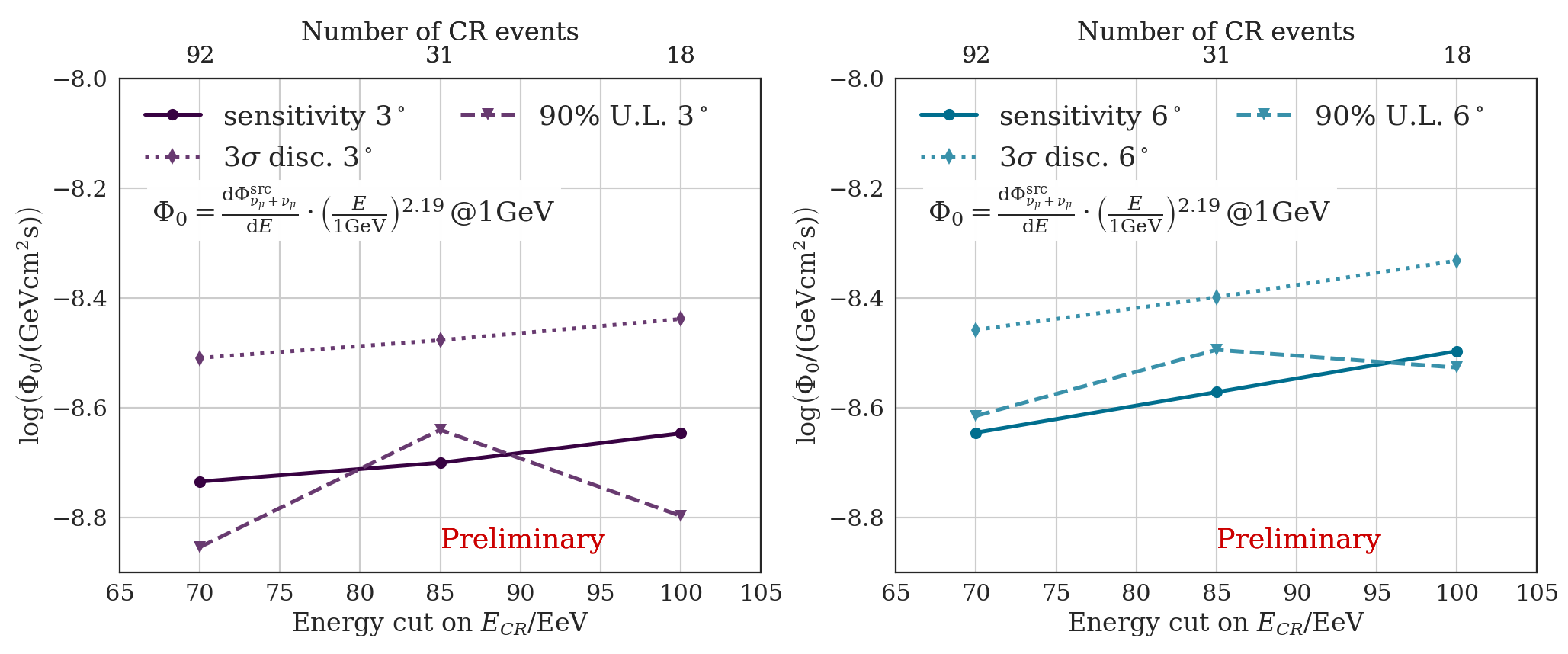}
  \caption{Experimentally obtained limits on the neutrino flux per source for each of the tested signal hypotheses, compared to sensitivity and $3\sigma$ discovery potential of the same signal model.}
  \label{fig:limits_result}
\end{figure}

The analysis is applied to the combined IC-ANT point-source data set.
Three different lower energy cuts $E_{CR}\geq[70~\mathrm{EeV}, 85~\mathrm{EeV}, 100~\mathrm{EeV}]$ are applied to the combined UHECR data set.
The magnetic deflection is scaled by $D=3^\circ$ and $6^\circ$.
Thus, six p-values are calculated w.r.t. an isotropic neutrino flux (Figure~\ref{fig:ts_result}).
All six test statistic values are well-compatible with the background expectation,
i.e. no significant clustering of neutrinos was found in the vicinity of UHECRs assuming a simple magnetic deflection described by the symmetric Gaussian smearing.
The results are used to calculate limits on the average neutrino flux per source (Figure~\ref{fig:limits_result}), assuming a signal model where every UHECR event has one neutrino source counterpart in its vicinity with a spectral index of $\gamma=2.19$.

\section{Discussion}
\label{sec:discussion}
The absence of spatial correlations between neutrinos and UHECRs has several explanations.
The two simplest ones are that UHECR sources do not produce a significant flux of neutrinos in the energy range where the present neutrino telescopes are sensitive or that the UHECR sources are transient such that the neutrino signal is not correlated on the time scale of the experiments.
However, if UHECRs and neutrinos do have common sources, it may be that either the UHECR 
deflections are too large to correctly estimate the source locations, or the neutrino flux is not strong enough to be detected by current experiments.
The former hypothesis could be investigated in the near future by including the UHECR composition and refined magnetic deflection estimates into the analysis, once such information is available.
The latter hypothesis is consistent with the background-compatible results of current time-integrated neutrino point-source searches~\cite{ic:reimann8yr,ant:9yrps} and strongly motivates Baikal-GVD, IceCube-Gen2 and KM3NeT in order to gather a larger amount of neutrino data.

\bibliography{template}

\end{document}